\documentclass[12pt,twoside]{article}
\usepackage[makeroom]{cancel}
\usepackage{latexsym}
\usepackage{longtable}
\usepackage{epsfig}
\usepackage{graphicx,bbm,psfrag}
\usepackage{amssymb,amsmath,amsthm,booktabs,mathtools}
\usepackage{bm}% bold math
\usepackage{color}

\newcommand{\bc}{\begin{center}}
\newcommand{\ec}{\end{center}}
\def\ba#1{\begin{array}{#1}\displaystyle}
\newcommand{\ea}{\end{array}}

\newcommand{\beq}{\begin{equation}}
\newcommand{\eeq}{\end{equation}}
\newcommand{\beqa}{\begin{eqnarray}}
\newcommand{\eeqa}{\end{eqnarray}}

\newcommand{\n}{\nonumber\\}
\newcommand{\bi}{\begin{itemize}}
\newcommand{\ei}{\end{itemize}}
\def\lt#1{\left#1}
\def\rt#1{\right#1}

\def\h#1{\hat{#1}}

\def\frc#1#2{\frac{#1}{#2}}

\newcommand{\p}{\partial}

\newcommand{\bra}{\langle}
\newcommand{\ket}{\rangle}

\newcommand{\R}{{\mathbb{R}}}

\newcommand{\ri}{{\rm i}}
\newcommand{\dd}{{\rm d}}

\def\eqref#1{(\ref{#1})}

\newcommand{\halmos}{\rule{1ex}{1.4ex}}
\newcommand{\eproof}{\hspace*{\fill}\mbox{$\halmos$}}

\usepackage{amsmath}	% required for `\align' (yatex added)
\begin{document}

\begin{center}
{\Large {\bf A geometric viewpoint on generalized hydrodynamics}}

\vspace{1cm}

{\large Benjamin Doyon$^*$, Herbert Spohn$^\dag$ and Takato Yoshimura$^*$}
\vspace{0.2cm}

{\small\em
$^*$ Department of Mathematics, King's College London, Strand, London WC2R 2LS, U.K.\\
$^\dag$ Physik Department and Zentrum Mathematik, Technische Universit\"at M\"unchen, Boltzmannstrasse 3, 85748 Garching, Germany}
\end{center}

\vspace{1cm}

\noindent Generalized hydrodynamics (GHD) is a large-scale theory for the dynamics of many-body integrable systems. It consists of an infinite set of conservation laws for quasi-particles traveling with effective (``dressed") velocities that depend on the local state. We show that these equations can be recast into a geometric dynamical problem.
They are conservation equations with state-independent quasi-particle velocities,  in a space equipped with a family of metrics, parametrized by the quasi-particles' type and speed, that depend on the local state. In the classical hard rod or soliton gas picture, these metrics measure the free length of space as perceived by quasi-particles; in the quantum picture, they weigh space with the density of states available to them. Using this geometric construction, we find a general solution to the initial value problem of GHD, in terms of a set of integral equations where time appears explicitly. These integral equations are solvable by iteration and provide an extremely efficient solution algorithm for GHD.
\vspace{1cm}

{\ }\hfill
\today

\section{Introduction}

Generalized hydrodynamics (GHD) \cite{CDY,bertini1} is a hydrodynamic theory
where the notion of local equilibration to a Galilean (or relativistic)
boost of a Gibbs state, is replaced by that of local relaxation to
generalized Gibbs ensembles \cite{GGE,eth2,EFreview}. It is expected to
emerge in appropriate hydrodynamic limits in both quantum and classical
integrable many-body systems, including field theory and spin chains,
and describes time evolution in inhomogeneous backgrounds or from
inhomogeneous states. It is in the context of steady states arising from
domain-wall initial conditions (see, for instance, a  review \cite{BDreview}) that it was originally
introduced, where it solved the long-standing problem of obtaining full
density and current profiles in interacting integrable quantum systems
\cite{CDY,bertini1}. It was generalized to include inhomogeneous force fields
\cite{force GHD}. It is seen as emerging in integrable classical systems
such as the hard rod fluid \cite{sphr,bds} and soliton gases \cite{solgas1,solgas2,solgas3,solgas4,dyc}, and a classical molecular dynamics solver has been developed for the general form of GHD \cite{dyc}. It was applied to study spin transport
\cite{spin1,spin2,spin3}, transport in the hard rod fluid \cite{ds}, and
quantum dynamics of density profiles such as propagating waves in interacting Bose gases \cite{bvkm2,ddky}. Most of these
studies concentrate on the emerging hydrodynamics at the Euler scale, at
which GHD was originally formulated, but see \cite{force GHD, prosendiff, ds} for discussions of viscosity effects. Here we do not discuss such effects.

The goal of this paper is twofold. First, we provide a geometric interpretation of GHD. We show that it is a theory for a gas of freely (inertially) propagating particles, but within a space whose metric depends both on the type and velocity of the particle, and on local distribution of particles in the gas. In the hard rod fluid, the metric  has a clear interpretation: it measures the free space available between the rods, a notion that was used in \cite{ds} in a derivation of the exact solution to the domain wall initial problem. The observation that this generalizes to soliton gases, still described by GHD, then suggests the metric construction proposed here. It is worth noting some similarity, in spirit, to Einstein's theory of general relativity, where currents are conserved in a metric that is determined by the matter content. Second, we use this geometric construction in order to provide a system of integral equations that {\em solve the initial-value problem of GHD} in full generality. The integral equations involve the initial condition and the time parameter in an explicit fashion, essentially integrating out the time direction. They can be solved by iteration. We confirm their validity by providing comparisons with direct solutions of the GHD partial differential equations. It is surprising that a general solution to a hydrodynamic equation can be obtained, and this might connect with the integrability of the GHD equations themselves, as found in soliton gases \cite{solgas1,solgas2,solgas3,solgas4,Bu17}.

The paper is organized as follows. In Section \ref{so} we review some of the main features of GHD. In Section \ref{sg} we develop the geometric interpretation of GHD in generality. In Section \ref{se}, we derive the integral equations that solve the initial value problem. Finally, in Section \ref{sc} we conclude, and in Appendix \ref{secg} we briefly explain the specialization to the hard rod problem.

\section{Overview of generalized hydrodynamics}\label{so}

The most powerful formulation of GHD to date \cite{CDY,bertini1} uses the physical notion of quasi-particles (although other formulations will doubtless come to the fore in the future). Integrable models solvable by Bethe ansatz, and integrable classical gases such as the hard rod model \cite{sphr,bds} and soliton gases \cite{dyc}, can be seen as models for interacting quasi-particles. A quasi-particle is specified by a spectral parameter $\theta\in\R$, parametrizing its momentum $p(\theta)$ (it can be taken as the velocity in the Galilean case, or the rapidity in the relativistic case; in general it is just a parametrization of the momentum). The scattering kernel $S(\theta,\alpha)$ characterizes the interaction amongst quasi-particles, and it is customary to define the differential scattering phase $\varphi(\theta,\alpha) = -\ri\,\dd \log S(\theta,\alpha)/\dd\theta$, which we assume to be symmetric. In this paper we assume for simplicity that there is a single particle species, but all equations are directly generalizable to models with many species such as the Heisenberg chain, simply by viewing the rapidity $\theta$ as a multi-index ${\bm\theta}=(\theta,j)$.

In the quasi-particle formulation of GHD, the local fluid state is described by a function specifying the distribution of quasi-particles. This function can be taken as the density $\rho_{\rm p}(x;\theta)$: the infinitesimal $\rho_{\rm p}(x;\theta)\,\dd x\,\dd \theta$ is the number of quasi-particles in the phase space volume element $[x,x+\dd x]\times [p(\theta),p(\theta) + p'(\theta)\dd\theta]$ (here and below the prime symbol ($'$) denotes a derivative with respect to the spectral parameter).

It is conventional to define the state density $\rho_{\rm s}(x;\theta)$ via
\beq
	 2\pi\rho_{\rm s}(x;\theta) = p'(\theta) + \int \dd \alpha\,\varphi(\theta-\alpha)\rho_{\rm p}(x;\alpha).
\eeq
In terms of these densities, one defines the occupation function
\beq\label{nrho}
	n(x;\theta) = \frc{\rho_{\rm p}(x;\theta)}{\rho_{\rm s}(x;\theta)}.
\eeq
The occupation function can be taken, instead of the quasi-particle density, as a characterization of the local fluid state. One may go from occupation function to quasi-particle and state densities as follows:
\beqa
	2\pi \rho_{\rm p}(x;\theta) &=& n(x;\theta) (p')^{\rm dr}_{[n(x)]}(\theta)\n
	2\pi \rho_{\rm s}(x;\theta) &=& (p')^{\rm dr}_{[n(x)]}(\theta)
	\label{eeq}
\eeqa
where the dressing operation depends on the function $n(x):\theta\mapsto n(x;\theta)$ (seen as an $x$-dependent function of the spectral parameter), and is defined, for any function $n$ (of the spectral parameter), by the solution to the following linear integral equation:
\beq
	h^{\rm dr}_{[n]}(\theta) = h(\theta) + \int \frc{\dd\alpha}{2\pi}
	\varphi(\theta,\alpha)n(\alpha)h^{\rm dr}_{[n]}(\alpha).
\eeq

As in any hydrodynamic theory, a model of generalized hydrodynamics also necessitates the equations of state: relations between the conserved currents and the conserved densities that emerge due to the interactions and dynamics of the constituents. This is determined by the energy function $E(\theta)$. More precisely, it was found in \cite{CDY,bertini1} that, in the quasi-particle picture of GHD, for each $\theta$, the function $\rho_{\rm p}(x;\theta)$ is a conserved density with associated current $v^{\rm eff}_{[n(x)]}(\theta)\rho_{\rm p}(x;\theta)$, where the effective velocity is given by \cite{CDY,bertini1,bonnes}
\beq\label{veff}
	v^{\rm eff}_{[n]}(\theta) = \frc{(E')^{\rm dr}_{[n]}(\theta)}{(p')^{\rm dr}_{[n]}(\theta)}.
\eeq
Equation \eqref{veff} can be seen as the equation of state.

The Euler-type hydrodynamic equations of GHD,  in the quasi-particle language, are therefore
\beq\label{cons}
	\p_t\rho_{\rm p}(x,t;\theta) + \p_x\big(v^{\rm eff}_{[n(x,t)]}(\theta)\rho_{\rm p}(x,t;\theta)\big)=0.
\eeq
These hold at the Euler scale, and thus omit viscosity terms; see
\cite{ds} for explicit viscosity terms in the hard rod fluid, and
\cite{force GHD} for an analysis of viscosity in integrable quantum models. Surprisingly, it turns out that the occupation function provides the normal modes of GHD, and thus is convectively conserved: Eq. \eqref{cons} is equivalent to \cite{CDY,bertini1}
\beq\label{neq}
	\p_t n(x,t;\theta) + v^{\rm eff}_{[n(x,t)]}(\theta)\, \p_x n(x,t;\theta) = 0.
\eeq
Note that these imply the continuity equation for the state density,
\begin{equation}\label{scont}
	\p_t\rho_{\rm s}(x,t;\theta) + \p_x\big(v^{\rm eff}_{[n(x,t)]}(\theta)\rho_{\rm s}(x,t;\theta)\big)=0. 
\end{equation}

The generalization to include force terms has also been obtained, see
\cite{force GHD}. In quantum field theory models, Eqs \eqref{cons} and \eqref{neq} are seen \cite{CDY}  to emerge from the conservation of local densities $\p_t\bra q_i(x,t)\ket + \p_x \bra j_i(x,t)\ket=0$, in the hydrodynamic approximation where averages are evaluated within entropy-maximized local states. A different argument  \cite{bertini1} based on a kinetic theory leads to the same equations in quantum chains.

The above formulation is valid in complete generality. However, there  are conditions on $\varphi(\theta,\alpha)$, $p(\theta)$, $E(\theta)$ and $\rho_{\rm p}(\theta)$ for it to provide physically sensible results. We do not know exactly what these conditions are, but the derivation presented below makes sense if the momentum derivative and the state density are strictly positive $p'(\theta)>0$, $\rho_{\rm s}(\theta)>0$.

\section{Geometry of GHD} \label{sg}

Eqs \eqref{veff} and \eqref{neq} can be viewed as arising from a  gas of colliding classical particles \cite{dyc}, generalizing the hard rod problem studied a long time ago \cite{bds}. Let us recall the main features of these models.

In the hard rod fluid, segments, all of fixed length $a$, move on the line, freely (inertially) except for collisions at which they exchange their velocities. By definition, a quasi-particle is a tracer of a given velocity: it follows the trajectory of a given velocity. The quasi-particle therefore travels like a free particle, except at collisions.  Two quasi-particles, with velocities $v,w$, collide if their actual positions, $x_w,x_v$ and velocities satisfy either $x_v - x_w = a$ and $v<w$, or  $x_w - x_v = a$ and $w<v$. Because of the collision, in the first case $x_v$ jumps to $x_v -a$ and $x_w$ to $x_w +a$, while in the second case $x_v$ jumps to $x_v +a$ and $x_w$ to $x_w -a$. The size of the jump is independent of $v,w$.

As a natural generalization, the jump size may be assumed to depend on the velocities of both collision partners. In other words, in the rules above, $-a$ is replaced by the general kernel $\varphi(v,w)$ without any particular sign restrictions. That is, $\varphi(v,w)$ is seen as an effective rod length, as perceived by the quasi-particles at velocities $v$ and $w$ when they collide. In order to take into account correctly cases where many particles interact within short time intervals, care must be taken in implementing the jumps. However, the above rules provide the sufficient features for our analysis, and for connecting with soliton scattering. See \cite{dyc} for more details. Heuristically, $\varphi(v,w)< 0$ corresponds to a repulsive and $\varphi(v,w)>0$ to an attractive interaction. In our context, velocity is replaced by the spectral parameter. Thus in a collision two quasi-particles at spectral parameters $\theta$ and $\alpha$ jump by the distance $-\varphi(\theta,\alpha)$ with a sign governed by the rules from above.

There is a natural geometric re-interpretation of the Euler equations for the hard rod or soliton gas. The idea is that the quasi-particles travel as free particles, which do not seem to interact, if one ``shrinks" the effective rods' lengths to zero. Indeed the motion of a test quasi-particle in the free space available in-between the (effective) rods in the gas in which it moves, is that of a free point-like particle at the group velocity $v^{\rm gr}(\theta) = E'(\theta)/p'(\theta)$. This free space is, however, state-dependent, as it depends on the density of rods in the fluid, and in the general case it also depends on the spectral parameters (the type and speed) of the test quasi-particle itself. Therefore, we may put a family of metrics on  the one-dimensional space in which quasi-particles propagate, parametrized by the spectral parameters, whose associated length is the available length as perceived by the test quasi-particles. Each such metric is determined by the local state, and in this metric, the test quasi-particle propagates freely.  The relation that fixes the metric as a function of the local state may be interpreted as an ``Einstein equation", and the free propagation is a conservation equation within the metric. The full system is therefore, in spirit, somewhat similar to Einstein's theory of general relativity.  

%%%%%%%%%%%%%%%%%%%%%%%%%%%%%%%%%%%%%%%%%%%%%%
\subsection{Metric and continuity equation}

In this subsection, we construct a dynamics on quasi-particles as per the above geometric arguments, and show that it gives rise to GHD.

Let us assume that there is some point $x_0$ such that the densities are independent of time $t$ for all $x\leq x_0,\; t\in[0,T]$. They are therefore also independent of $x$ in this region by the continuity equation:
\beq\label{cond1}
	\rho_{\rm p}(x,t;\theta) = \rho_{\rm p}^-(\theta)\quad\mbox{for all}\quad x\leq x_0,\; t\in[0,T],\;\theta\in\R.
\eeq
Then the derivation below will be valid for all times up to $T$.  The choice of the left side is purely conventional. We will use the symbol $n^-(\theta)$ for the associated occupation function, and $\rho_{\rm s}^-(\theta)$ for the state density.

This assumption is immediately satisfied if all particles are initially located in a bounded region of phase space, as well as for domain-wall initial conditions \cite{BDreview,CDY,bertini1}, with baths that extend both on the left and the right (in both cases the assumption holds for every finite $T$ by choosing $|x_0|$ large enough).

Let us consider the phase space $(x;\theta)\in\R\times \R$. We put on this space the volume element $\dd V$ equal to the ``available", or ``free", volume: the space available within $\dd x$ as seen by a test quasi-particle at rapidity $\theta$, times $2\pi \dd \theta$:
\beq
	\dd V= \rho_{\rm s}(x;\theta)\, \dd x \,2\pi \dd \theta.
\eeq
Convenient coordinates are the $(q,p)$ coordinates, where $p$ is the momentum and the volume element is $\dd V = \dd q \,\dd p$. This volume element therefore gives a relation between $\dd q$ and $\dd x$, inducing a one-dimensional, $\theta$- and state-dependent metric on $x$-space,
\beq\label{dqdx}
	\dd q = \mathcal{K}_{[n(x)]}(\theta)\,\dd x.
\eeq
One takes $\dd q^2 = g(x;\theta)\,\dd x^2$ as the infinitesimal square length at constant $\theta$ with metric $g(x;\theta) = \mathcal{K}_{[n(x)]}(\theta)^2$, and
\beq\label{K}
	\mathcal{K}_{[n(x)]}(\theta)= \frc{2\pi}{p'(\theta)}\rho_{\rm s}(x;\theta)  = \frc{(p')^{\rm dr}_{[n(x)]}(\theta)}{p'(\theta)}.
\eeq
The quantity $\dd q$ may be interpreted as the infinitesimal number of states available within the interval $[x,x+\dd x]$ per unit momentum, as seen by a test quasi-particle at spectral parameter $\theta$. In this point of view, there is a one-parameter family of metrics, in bijection with the one-parameter family of conserved quasi-particle numbers $n(\theta)$. Each metric relates to the one-dimensional space on which the quasi-particle at spectral parameter $\theta$ moves. Clearly, these metrics depend on the local state. The coordinate $q$ is defined by integrating from $x_0$:
\beq
	q = q(x;\theta) = \frc{2\pi }{p'(\theta)}\int_{x_0}^x \dd y\,\rho_{\rm s}(y;\theta).
\eeq

Consider the free density of particles $n(x;\theta)$, the density per unit free volume $V$. The total number of particles within $\dd V$ is
\beq
	n(x;\theta)\, \dd V = \rho_\mathrm{p}(x;\theta)\,\dd x \,2\pi \dd \theta
\eeq
and therefore
\beq
	n(x;\theta) =  \frc{\rho_{\rm p}(x;\theta)}{\rho_{\rm s}(x;\theta)}.
\eeq
We thus see that the occupation function \eqref{neq} is equal to the particle density per unit free volume in the geometric language. We define $\h n(q;\theta)$ by $\h n(q(x;\theta);\theta) = n(x;\theta)$.

We now put a dynamics on this system, evolving in time $t$. All quantities acquire a $t$ dependence, including the relation between the $q$ coordinate and the $x$ coordinate,
\beq\label{qxt}
	q(x,t;\theta) = \frc{2\pi }{p'(\theta)}\int_{x_0}^x \dd y\,\rho_{\rm s}(y,t;\theta).
\eeq
The quasi-particles are deemed to propagate ballistically in the free space. From the above construction, it seems natural that the velocity of propagation depend on the asymptotic particle density $\rho_{\rm p}^-(\theta)$. We choose the following definition of the propagation velocity:
\beq\label{vmoins}
	v^-(\theta) = v^{\rm eff}_{[n^-]}(\theta){\cal K}_{[n^-]}(\theta) = \frc{(E')^{\rm dr}_{[n^-]}(\theta)}{p'(\theta)}.
\eeq
This has the interpretation as the asymptotic effective velocity as expressed with respect to $q$-space, thus multiplied by the square-root of the metric. The fluid variable $\h n(q,t;\theta)$ therefore satisfies the equation for free particle motions at the velocity $v^-(\theta)$ as function of the position in the free space $q$ and of time $t$:
\beq\label{free}
	\p_t \h n(q,t;\theta) + v^-(\theta)\, \p_q \h n(q,t;\theta)=0
\eeq
where $\h n(q,t;\theta)$ is defined by
\beq\label{hn}
	\h n(q(x,t;\theta),t;\theta) = n(x,t;\theta).
\eeq
We now show that \eqref{free} is equivalent to \eqref{neq}.

\proof The proof we present goes in one direction of the equivalence, the other direction being immediate. From \eqref{hn}, we have
\beq
	\p_t n(x,t;\theta) = \p_t \h n(q,t;\theta) + \p_t q(x,t;\theta)\,
	\p_q \h n(q,t;\theta).
\eeq
Therefore, using \eqref{free}, as well as \eqref{dqdx} and \eqref{K} in order to relate $\p_q$ to $\p_x$, we obtain
\beq\label{eq1}
	\p_t n(x,t;\theta)  + \frc{p'(\theta)\,(\h v^-(\theta) - \p_t q(x,t;\theta))}{2\pi \rho_{\rm s}(x,t;\theta)}
	\p_x n(x,t;\theta)=0.
\eeq
We evaluate $\p_t q(x,t;\theta)$ by using \eqref{scont}:
\beqa
	\p_t q(x,t;\theta) &=& \frc{2\pi}{p'(\theta)}
	\int_{x_0}^x \dd y\,\p_t\rho_{\rm s}(y,t;\theta) \n
	&=& -\frc{2\pi}{p'(\theta)}
	\int_{x_0}^x \dd y\,\p_y\Big(v^{\rm eff}_{[n(y,t)]}(\theta)
	\rho_{\rm s}(y,t;\theta)\Big) \n
	&=& -\frc{2\pi}{p'(\theta)}
	\Big(v^{\rm eff}_{[n(x,t)]}(\theta)
	\rho_{\rm s}(x,t;\theta) - v^{\rm eff}_{[n^-]}(\theta)
	\rho_{\rm s}^-(\theta)\Big) \n
	&=& -\frc{2\pi}{p'(\theta)}
	v^{\rm eff}_{[n(x,t)]}(\theta)
	\rho_{\rm s}(x,t;\theta) + v^-(\theta)\label{abf}
\eeqa
from which \eqref{neq} follows.
\eproof

If the asymptotic particle density is zero, then $v^-(\theta) = v^{\rm gr}(\theta)$ is the group velocity: in this case, the particles in $q$ space propagate as free particles at the group velocity. The derivation above is unaffected by the change $\theta \mapsto (\theta,j$) where $j$ is an internal index for particle types; the only difference, in the derivation, is that integrals over spectral parameters are augmented by sums over particle types. Therefore the result stays the same in models with many particle species. The set of  ``Einstein-Euler" equations \eqref{qxt} and \eqref{free} can be interpreted as an infinite set of Euler-type conservation laws, one for each quasi-particle spectral parameter, in one-dimensional spaces whose metrics, one for each spectral parameter, are determined by the local state.

\subsection{GHD and invariance of volume form}
The choice of the metric \eqref{K} is also natural if we consider GHD
as a {\it non-Hamiltonian system} \cite{nonh}, which is a classical system whose
determinant of the Jacobian associated with the coordinate
transformation $(x_0,p_0)\mapsto (x_t,p_t)$ is 
not unity: the standard volume element $\dd
x\,\dd p$ is not preserved in the course of time-evolution. Volume preservation is recovered under an appropriate choice of metric. Here, we show that preservation of the volume element $\dd q\wedge\dd p = \mathcal{K}_{[n(x,t)]}(\theta) \,\dd x\wedge\dd p$ along the path $(\dot{x}_t,\dot{p}_t)=(v^{\rm eff}_{[n(x,t)]}(\theta),0)$ of a test particle within GHD, amounts to the continuity equation \eqref{scont} for $\rho_{\rm s}(x,t;\theta)$

Let us denote the standard volume form as a differential 2-form $\dd x_t \wedge
\dd p_t$. Then,
\begin{align}
  \frac{\dd}{\dd t}\dd V&=\lt(\frac{\dd}{\dd t}\mathcal{K}_{[n(x,t)]}(\theta)\rt)\dd
 x_t\wedge\dd p_t+\mathcal{K}_{[n(x,t)]}(\theta)\lt(\frac{\dd}{\dd t}\dd x_t\rt)\wedge\dd
 p_t+\mathcal{K}_{[n(x,t)]}(\theta)\dd x_t\wedge\lt(\frac{\dd}{\dd t}\dd p_t\rt) \n
&=\bigl[\p_t\mathcal{K}_{[n(x,t)]}(\theta)+v^{\rm
 eff}_{[n(x,t)]}(\theta)\p_x\mathcal{K}(\theta)\bigr]\dd
 x_t\wedge \dd p_t+\mathcal{K}_{[n(x,t)]}(\theta)\p_xv^{\rm
 eff}_{[n(x,t)]}\dd x_t\wedge \dd p_t, \n
&=\bigl[\p_t\mathcal{K}_{[n(x,t)]}(\theta)+\p_x(v^{\rm
 eff}_{[n(x,t)]}(\theta)\mathcal{K}_{[n(x,t)]}(\theta))\bigr]\dd
 x_t\wedge \dd p_t, \label{Kcont}
\end{align}
where in the second line we used 
\begin{equation}
\lt(\frac{\dd}{\dd t}\dd x_t\rt)\wedge\dd p_t=\dd \dot{x}_t\wedge\dd p_t=(\p_xv^{\rm
  eff}_{[n(x,t)]}(\theta)\dd x_t+\p_p v^{\rm eff}_{[n(x,t)]}(\theta)\dd p_t)\wedge\dd
  p_t=\p_xv^{\rm eff}_{[n(x,t)]}\dd x_t\wedge \dd p_t 
\end{equation}
and $\dd x_t\wedge\lt(\frac{\dd}{\dd t}\dd p_t\rt)=0$. Thus $\frac{\dd}{\dd
t}\dd V_t=0$ implies \eqref{scont}.

Likewise, the invariance of the density element $n(x,t;\theta)\dd V_t$
under the time-evolution leads to the continuity equation
\eqref{cons}. This is a natural generalization of the Liouville theorem
to the non-Hamiltonian systems. It is also a
simple matter to extend this derivation to the case where a system is
subject to an external force.

\section{Integral equations solving the initial value problem} \label{se}

We have shown that the set of equations \eqref{qxt}, \eqref{free} and \eqref{hn} is equivalent to \eqref{neq}. Therefore, we may look for solving these simpler-looking equations.  This turns out to be possible, providing a scheme that is a very efficient numerical method for solving the GHD initial value problem for a time evolution without inhomogeneous external force field.

\subsection{Solution to the initial value problem}

Let $n^0(x,\theta)$ be the initial occupation function, with associated state density $\rho_{\rm s}^0(x,\theta)$. Recall the metric \eqref{K}. Clearly, if there is no interaction, $\varphi(\theta,\alpha)=0$, then $\mathcal{K}_{[n]}(\theta) = 1$. We show that the solution to the initial value problem for \eqref{neq} is given by
\beq\label{soln}
	n(x,t;\theta) = n^0(u(x,t;\theta);\theta)
\eeq
where $u(x,t;\theta)$ is determined implicitly by the integral equation
\beq\label{y}
	\int_{x_0}^{u(x,t;\theta)} \dd y\,\mathcal{K}_{[n^0(y)]}(\theta) + v^-(\theta)\, t = \int_{x_0}^x \dd y\,\mathcal{K}_{[n(y,t)]}(\theta).
\eeq
These simply indicate that the particles move at the velocity $v^-(\theta)$ in the free space. Note that $u(x,t;\theta)$ is monotonically increasing as a function of $x$ because $\mathcal{K}_{[n]}(\theta)$ is positive. Equations \eqref{soln} and \eqref{y} are derived as follows. Solving \eqref{free} is simple:
\beq
	\h n(q,t;\theta) = \h n^0(q-v^-(\theta)t;\theta)
\eeq
where $\h n^0(q(u,0;\theta),\theta) = n^0(u;\theta)$ is the initial condition in the $q$-space. Therefore $n(x,t;\theta) = \h n^0(q(x,t;\theta)-v^-(\theta)t;\theta)$. Hence we must define $u=u(x,t;\theta)$ as solving
\beq\label{dxr}
	q(u(x,t;\theta),0;\theta) = q(x,t;\theta)-v^-(\theta)t.
\eeq
This immediately yields \eqref{y}.

The explicit time parameter appears in \eqref{y}. Given the initial fluid states at every point $y$ we may construct $\mathcal{K}_{[n_0(y)]}$. With this, we then must solve simultaneously the system of integral equations \eqref{soln} and \eqref{y} for the unknowns $n(x,t;\theta)$ and $u(x,t;\theta)$ as functions of $(x;\theta)$ for $t$ fixed. Here \eqref{soln} involves $n(x,t;\theta)$ and $u(x,t;\theta)$, and \eqref{y} involves $u(x,t;\theta)$, and involves $n(y,t;\alpha)$ for all $y\in[x_0,x]$ and all $\alpha\in\R$. In the case without interaction, we immediately find the expected relation $u(x,t;\theta) = x-v^-(\theta) t$ with $v^-(\theta) = v^{\rm gr}(\theta)$. In general, $u(x,t;\theta)$ satisfies the differential equation
\beq
	\mathcal{K}_{[n^0(u(x,t;\theta))]}(\theta)\, \p_tu(x,t;\theta)
	+\mathcal{K}_{[n(x,t)]}(\theta)\,v^{\rm eff}_{[n(x,t)]}(\theta)\p_x u(x,t;\theta) = 0
\eeq
and, using \eqref{soln}, this can be seen as a generalization to GHD of the
solution of a single-component fluid by the method of characteristics. In particular, in a uniform, stationary state where $n(x,t;\theta)=n(\theta)$ is independent of $x$ and $t$, we have $u(x,t;\theta) = x-v^{\rm eff}_{[n]}(\theta)\, t$, representing the propagation of a particle at rapidity $\theta$ within a state characterized by $n$.

In the case of the domain-wall initial condition, this general solution specializes to that provided in \cite{CDY,bertini1,ds}. Indeed, assume that $n^0(x;\theta)=n^-(\theta)$ for $x<0$,  and $n^0(x;\theta)=n^+(\theta)$ for $x>0$. Since $u(x,t;\theta)$ is monotonic with $x$, we must find $x_*=x_*(t;\theta)$ solving $u(x_*,t;\theta)=0$; we have $n(x,t;\theta) = n^-(\theta)\chi(x<x_*(t;\theta)) + n^+(\theta)\chi(x>x_*(t;\theta))$ where $\chi$ is the indicator function. By \eqref{dxr}, differentiating with respect to $t$ and taking into account \eqref{qxt}, this implies
\beq
	\p_t q(x_*,t;\theta) - v^-(\theta)
	+\p_t x_*(t;\theta)\,\frc{2\pi}{p'(\theta)}\rho_{\rm s}(x_*,t;\theta) (\theta) = 0.
\eeq
From the result \eqref{abf} this gives
\beq
	\p_t x_*(t;\theta) = v^{\rm eff}_{[n(x_*(t;\theta),t)]}(\theta).
\eeq
With scale-invariant initial condition, it is natural to assume that the solution to the scale-invariant equations \eqref{neq} is self-similar (that is, depends on the ray $x/t$ only). If we set $x_*(t;\theta) = t\xi_*(\theta)$, then the solution presented here is self-similar, and $n(x_*(t;\theta),t) = n(\xi_*(\theta),1)$. In this case we find
\beq
	n(x,t;\theta) = n^-(\theta)\chi(\xi<\xi_*(\theta)) + n^+(\theta)\chi(\xi>\xi_*(\theta)),\quad \xi_*(\theta) = v^{\rm eff}_{[n(\xi_*(\theta),1)]}(\theta).
\eeq
which is indeed the solution given in \cite{CDY,bertini1,ds}.

We do not know yet how to address the uniqueness of the solution to \eqref{soln}, \eqref{y}, and in particular how to show that self-similarity must hold in the domain-wall problem. However, we note that, in the hard rod problem, the idea of ``shrinking" the rods to points and then freely evolving them in time, which directly corresponds to the integral equations \eqref{soln}, \eqref{y}, was used in \cite{bds} in order to show uniqueness properties. Thus the above may give in the future some insight into this problem.

Finally, again, it is clear that all the above results hold under $\theta \mapsto (\theta,j)$ when many particle types are involved.

\subsection{Numerical method}

The set of equations \eqref{soln}, \eqref{y} are of interest from a theoretical perspective. But at the same time they furnish an efficient algorithm for numerically solving the initial value problem of GHD.

In general, Equations \eqref{soln}, \eqref{y} can be solved by the following iteration scheme. First, one sets, as an initial condition to the iteration scheme, $n(x,t;\theta) = n_0(x;\theta)$. Note that with the knowledge of $n(x,t;\theta)$ and $n_0(x;\theta)$, both integrands in \eqref{y} can be evaluated. One then solves \eqref{y} for $u(x,t;\theta)$, and constructs a new iteration of $n(x,t;\theta)$ using \eqref{soln}. The process is then repeated until $n(x,t;\theta)$ converges. Alternatively, one may set, as an initial condition to the iteration scheme, $n(x,t;\theta)$ to the right-hand side of \eqref{soln} with $u(x,t;\theta) = x-v^-(\theta)t$. The latter only depends on the asymptotic velocity $v^-(\theta)$ \eqref{vmoins}, whose evaluation only requires the knowledge of the initial state.

In order to explicitly check that solving the integral
equations by iteration coincides with directly solving the GHD equation
\eqref{neq}, we focus on the ``bump-release problem" studied for instance in \cite{ddky}. In this problem, a Lieb-Liniger (LL) gas, with $\varphi(\theta,\alpha) =
2c/((\theta-\alpha)^2+c^2)$, is initially in the ground state or a finite-temperature state of a potential with an inverted Gaussian centered at the origin, where a density accumulation occurs. The potential is then suddenly flattened, so that the density accumulation is released into two oppositely propagating waves. At zero temperature, GHD results have been explicitly checked in \cite{ddky} to agree with numerical quantum evolution, and it has been shown that GHD equations \eqref{neq} simplify to conventional hydrodynamics for finite times, providing a simple alternative for the time evolution. This allows for compelling comparison with the current iterative method. We also study the bump release problem at finite temperature where conventional hydrodynamics fails, comparing with the molecular dynamics developed in \cite{dyc} and thus giving further compelling evidence.
\begin{figure}[h]
	\centering
	\includegraphics[width=14.0cm]{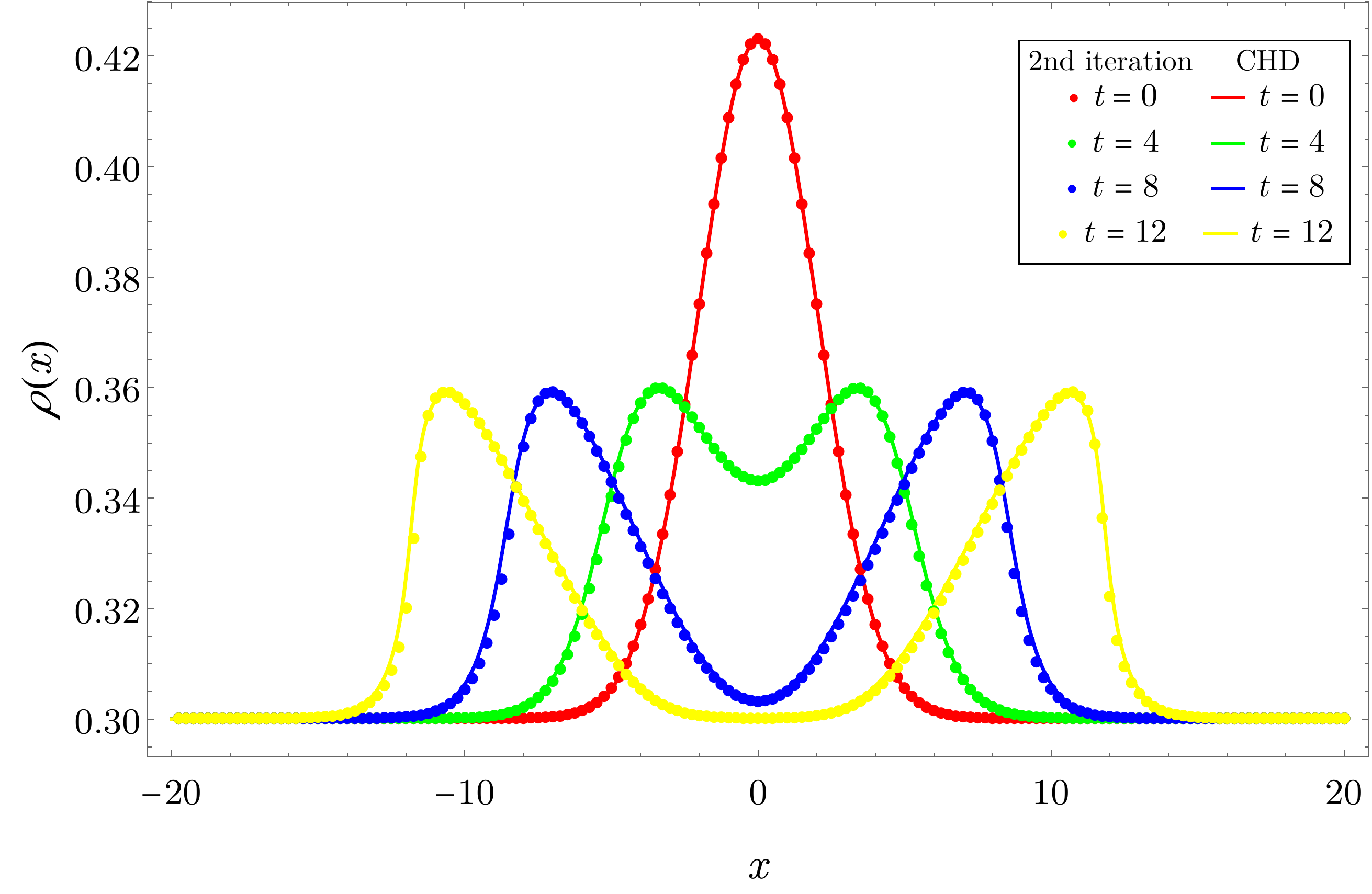}
	\caption{Evolution with $c=0.5$ and initial potential
	 $V(x)=-0.1e^{-x^2/8}-\mu_\infty$. The background chemical potential $\mu_\infty$ is such
	 that $\rho_{\infty}=0.512$. The simple codes have been implemented in the Mathematica software, and use the interpolation functionality for improving the accuracy of the finite-element approximation of integrals. Points are from the second iteration of \eqref{soln}, \eqref{y}, and are separated by the actual value of $\Delta x$ used for the finite-element approximation. Solid lines are from directly solving conventional hydrodynamics (CHD), shown to be equivalent to GHD in \cite{ddky}.}
\label{f1}
\end{figure}
\begin{figure}[h]
	\centering
	(a) \hspace{-1cm} \includegraphics[width=7.5cm]{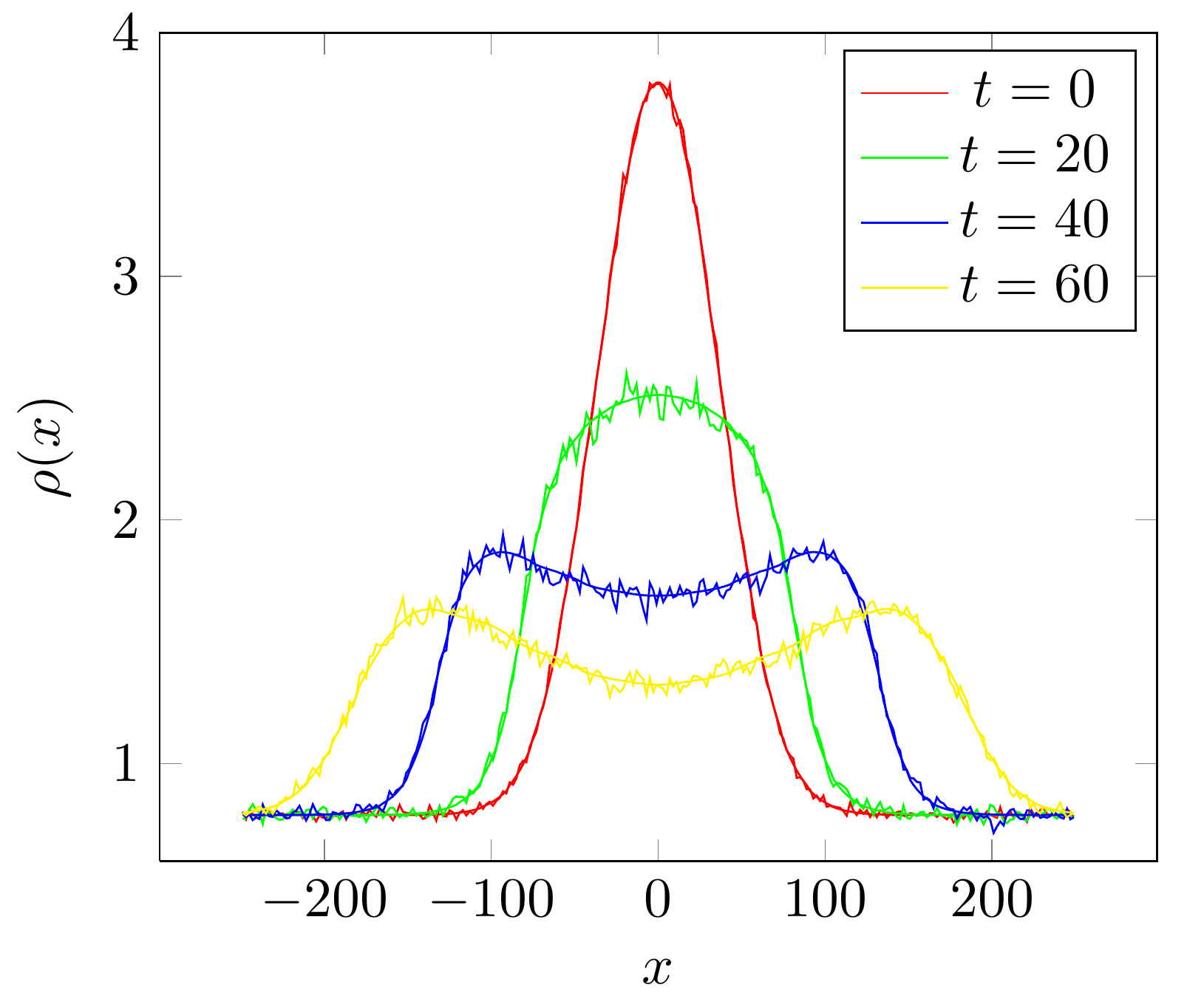}
	\quad 
	(b) \hspace{-1cm}\includegraphics[width=7.5cm]{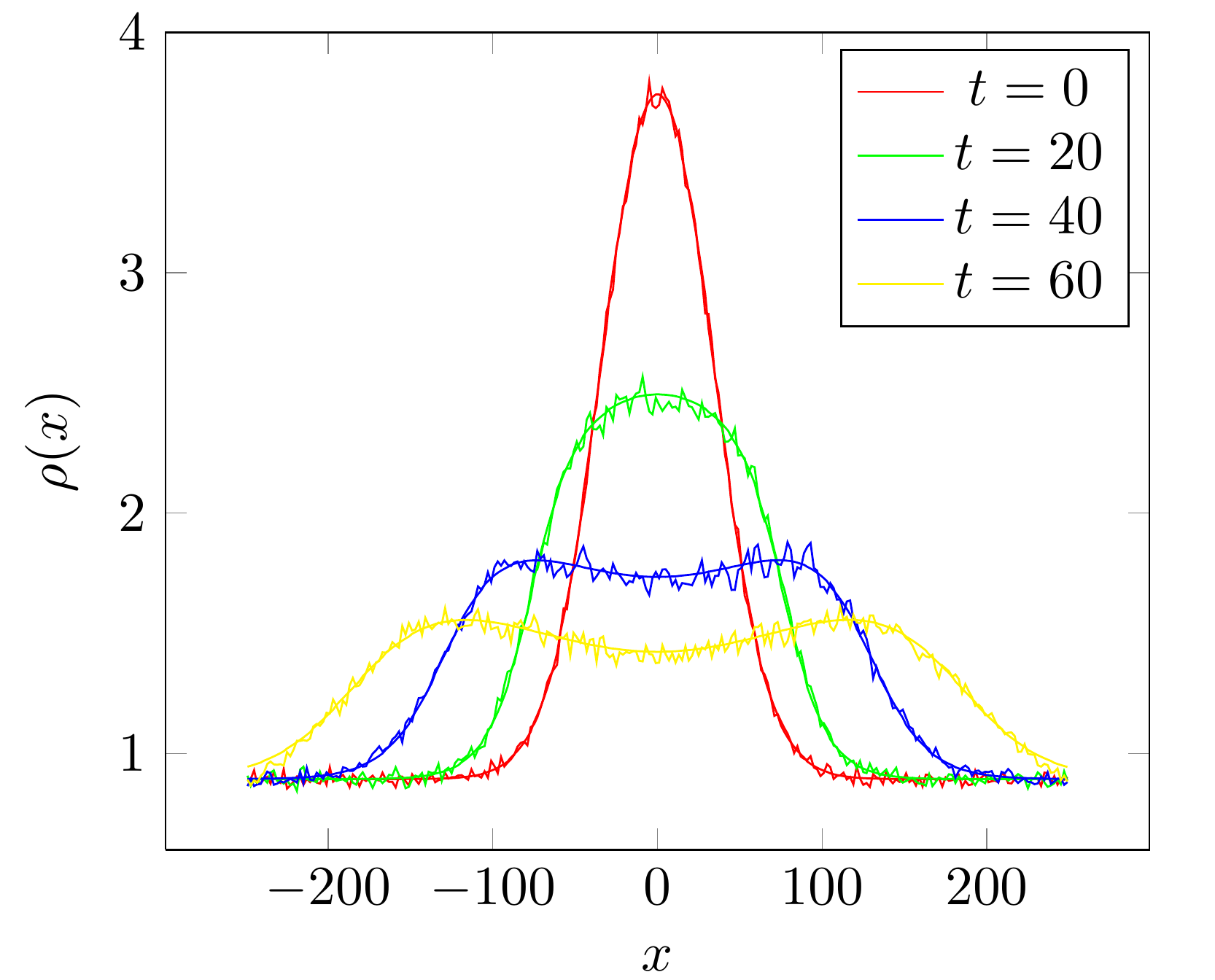}
	\caption{Evolution with $c=2$, from in initial states with potential
	 $V(x)= -  5 e^{ - (x/50)^2} -1$ at temperatures $T=1$ (a) and $T=2$ (b). Codes have been implemented in the c programming language. Smooth curves are from the fourth iteration of \eqref{soln}, \eqref{y}, using 800 divisions along $x\in[-375,375]$, and 800 divisions along $\theta\in[-7,7]$ for finite-element approximations. The superimposed noisy curves are from the molecular dynamics (the flea gas algorithm) introduced in \cite{dyc}, with around 860 (a) and 910 (b) particles and 2000 samples.}
\label{f2}
\end{figure}

Quite surprisingly, we observe that a few iterations, of the order of 2 to 4, are enough to
accurately  reproduce the space-time profile of the density $\rho(x,t) = \int \dd \theta\, \rho_{\rm p}(x,t;\theta)$.

In Fig. \ref{f1}, we depict the time-evolution of the LL gas with a
coupling constant $c=0.5$ (with mass set to 1) upon a release from the ground state of an energy potential of the form $V(x)=-0.1e^{-x^2/8}-\mu_\infty$ (coupling to the density of particles). This is a perturbation of a background chemical potential $\mu_{\infty}$ corresponding to the background density
$\rho_{\infty}=0.512$. The initial fluid state is obtained by a local density approximation of the ground state. With this choice of parameters, the
interaction is strong enough to render the dynamics nontrivial. The solid lines represent curves from a direct solution of conventional hydrodynamics using finite-element approximations, and dots are from the second iteration of \eqref{soln} and \eqref{y}. Agreement is excellent. We show results only for times before the formation of sharp structures, where conventional hydrodynamics fails (these are ``dissolving shock," and the solution beyond this time was first obtained in \cite{ddky}). Naturally, the steep wave front that develops would necessitate a smaller finite element $\Delta x$ in order to improve precision. 

A similar observation is made at finite temperature. In Fig. \ref{f2}, we depict the time-evolution of the LL gas with $c=2$, and initial state obtained by a local density approximation of the potential $V(x)= -5\, e^{-(x/50)^2} -1$ at temperatures $T=1$ (a) and $T=2$ (b). We used 4 iterations, and enough finite-element divisions for high precision results. It required about two minutes of computer time on a standard laptop for every of the four curves in each graph. Using the iterative method with $n_0(x;\theta)$ set to the $t=0$ initial value, we did not see any degradation in precision as a function of time $t$. Comparison is made against a simulation using molecular dynamics (the flea gas algorithm) \cite{dyc}, with excellent agreement. There are only small differences between the cases $T=1$ and $T=2$, which seem to be well captured by the iterative method. The iterative method is much more precise than molecular dynamics, for less computer time.

Other algorithms for solving GHD are (1) the approximate scheme used in \cite{bvkm2}, essentially based on an approximation of the present algorithm that is valid for small times; (2) the direct solution to the finite hydrodynamics to which GHD reduces in the zero-entropy subspace \cite{ddky}; and (3) the molecular dynamics of \cite{dyc}. The latter two are the only ones valid also for evolution within external, inhomogeneous force fields. However,  without such force fields, the numerical algorithm for the GHD initial value problem developed here appears to be very efficient, providing accurate solutions for the full distribution $n(x,t;\theta)$, at any given time $t$, within at most few minutes of laptop computer time.

%%%%%%%%%%%%%%%%%%%%%%%%%%%%%%%%%%%%%%%%%%%
\section{Conclusion}\label{sc}

We have shown that the equations of GHD (those valid at the Euler scale), in the quasi-particle formulation, can be recast into hydrodynamic conservation equations for a gas of particles freely propagating at velocities  that are determined by the homogeneous and stationary state asymptotically far in space. This gas propagates in a space whose metric is proportional to the local state density. In the classical interpretation of GHD, it measures the effective space available for quasi-particles to travel between collisions. In the quantum interpretation, it measures distances by weighing with the number of Bethe roots available in local states, as if the actual space in which a particle propagates were proportional to the ``Bethe space" \footnote{We thank Bruno Bertini for suggesting this interpretation.}. The structure is somewhat similar to that of Einstein's equations for general relativity, where a metric is dynamical, determined by the particle content at that space-time point, and particles satisfy conservation equations within this metric. In the present case, because of the infinity of conservation laws, there are infinitely many conserved quasi-particle numbers, parametrized by the spectral parameter $\theta$, and there is a one-parameter family of metrics, as perceived by test quasi-particles at spectral parameters $\theta$.

Importantly, we have shown that this new viewpoint allows for an exact solution to the initial value problem of GHD. The exact solution is expressed as a system of nonlinear integral equations, Equations \eqref{soln} and \eqref{y}, which are readily solvable on standard laptop computer.

This geometric construction seems to connect naturally with the recent work \cite{dubail} showing how inhomogeneous states and evolutions in free fermion models can be recast into field theory on curved space. It would be interesting to analyze the uniqueness properties of solutions to Equations \eqref{soln}, \eqref{y}. Using these equations, the approach to stationary states could also be analyzed. It would also be interesting to generalize the above geometric construction to the presence of force fields. Finally, there is an elegant symplectic structure underlying the above geometric construction, with symplectic form $\dd q\wedge \dd p$, which we plan to develop in the future.

\medskip

{\bf Acknowledgments.} We thank B. Bertini for discussions and J. Dubail for insightful comments on the manuscript. BD and TY thank City University New York (workshop ``Dynamics and hydrodynamics of certain quantum matter", March 2017) for hospitaliy. TY is grateful for the support from the Takenaka Scholarship Foundation.

\medskip
{\em Note added:} As the first version of this paper was in its last stage of preparation, the work \cite{bvkm2} appeared which, amongst other things, develops aspects of the semi-Hamiltonian and integrability structures of GHD, see also \cite{Bu17}. A natural geometry appears within this context, with, surprisingly, a metric of similar form to that introduced here, and a certain formal solution by quadrature. It would be very interesting to understand the relation between these two viewpoints.

\appendix

\section{The hard rod fluid}\label{secg}

The geometric interpretation developed in the main text can be specialized to the hard rod fluid; it is instructive to see this explicitly. The system is Galilean with a single quasi-particle specie, and we choose unit mass. Let us use the more standard notation $v=\theta$ for the velocity, and choose $p(v)=v$ and $\varphi(v,w)=-a$, where $a$ is the length of the rods.

The particle density is $\rho(x) = \int \dd v\rho_{\rm p}(x;v)$. The density of state simplifies and is independent of $v$, with $2\pi \rho_{\rm s}(x) = 1-a\rho(x)$.  We denote the average velocity by
\beq
	u(x) = \rho(x)^{-1}\int \dd v\,\rho_{\rm p}(x;v).
\eeq
The effective velocity was shown in \cite{ds} to be $v^{\rm eff}_{[n]}(v) = (v-a\rho u)/(1-a\rho)$. Similarly to what we did in the main text, we assume that there is some point $x_0$ such that the densities on its left are all independent of space-time $t$ up to some time $T$.

The geometry is very clear in this case, as the coordinate $q$ corresponds to shrinking the rod length $a$ to zero. The length element $\dd q$ is related to the original coordinate's infinitesimal $\dd x$ as
\beq
	\dd q = (1-a\rho(x))\dd x
\eeq
with $0 \leq a \rho(x) \leq 1$.  Contrary to the general case, it does not depend on the velocity. The density of particle with respect to the new metric is
\beq
	n(x;v) = \frac{ \rho_{\rm p}(x;v)}{1-a\rho(x)},
\eeq
which was referred to as the free density in \cite{ds} (the specialization of \eqref{nrho} to the hard rod case). In the $q$ coordinate, taking into account the asymptotic bath with density $\rho^-$ and average velocity $u^-$, particles move at velocities $v^-(v) = v-a\rho^- u^-$. We define
\beq\label{defs}
	q = q(x,t) =  \int^x_{x_0} \dd x'\, (1-a\rho(x'))
\eeq
and $\h n(q,t;v)$ satisfying $\h n(q(x,t),t;v) = n(x,t;v)$. As in the main text, one can show that the trivial evolution
\beq\label{freerods}
	\p_t \h n(q,t;v) + v^-(v) \,\p_q \h n(q,t;v) = 0
\eeq
is equivalent to $\p_t n(x,t;v) + v^{\rm eff}_{[n(x,t)]}(v)\p_x n(x,t;v) = 0$.

The exact solution to the initial value problem simplifies thanks to the lack of $v$ dependence of the state density. Let $n^0(x;v)$ be the initial free density, with associated total density $\rho^0(x)$. The solution is
\beq\label{solnrods}
	n(x,t;v) = n^0(y;v)
\eeq
where $y=y(x,t;v)$ is determined implicitly by the equation
\beq\label{yrods}
	\int_{x_0}^y \dd y'\,(1-a\rho^0(y')) + v^-(v)t = \int_{x_0}^x \dd x'\,(1-a\rho(x',t)).
\eeq


\begin{thebibliography}{99}
 \bibitem{CDY} O. A. Castro-Alvaredo, B. Doyon and T. Yoshimura, ``Emergent hydrodynamics in integrable quantum systems out of equilibrium
'', Phys. Rev. X {\bf 6}, 041065 (2016).
\bibitem{bertini1} B. Bertini, M. Collura, J. De Nardis and M. Fagotti, ``Transport in out-of-equilibrium XXZ chains: exact profiles of charges and currents'', Phys. Rev. Lett. {\bf 117}, 207201 (2016).
\bibitem{GGE} M. Rigol, V. Dunjko, V. Yurovsky and M. Olshanii,
	``Relaxation in a Completely Integrable Many-Body Quantum
	System: An Ab Initio Study of the Dynamics of the Highly Excited
	States of 1D Lattice Hard-Core Bosons''. Phys. Rev. Lett. {\bf 97},
	050405 (2007).
\bibitem{eth2} J. Eisert, M. Friesdorf and C. Gogolin, ``Quantum
	many-body systems out of equilibrium'', Nat. Phys. {\bf 11}, 124
(2015).

\bibitem{EFreview} F. Essler and M. Fagotti, ``Quench dynamics and
	relaxation in isolated integrable quantum spin chains'', J. Stat. Mech. \textbf{2016}, 064002 (2016), special issue on {\em Nonequilibrium
	dynamics in integrable quantum systems}.
\bibitem{BDreview} D. Bernard and B. Doyon, ``Conformal field theory out of
equilibrium: a review", J. Stat. Mech. {\bf 2016}, 064005
(2016).
\bibitem{force GHD} B. Doyon and T. Yoshimura, ``A note on generalized
	hydrodynamics: inhomogeneous fields and other concepts'', SciPost Phys. {\bf 2}, 014 (2017).
\bibitem{sphr}  H. Spohn, ``Hydrodynamical theory for equilibrium time correlation functions of hard rods",  Annals of Physics {\bf 141}, 353 (1982).
\bibitem{bds} C. Boldrighini, R. L. Dobrushin and Yu. M. Sukhov, ``One-dimensional hard rod caricature of hydrodynamics", J. Stat. Phys. {\bf 31}, 577 (1983).

\bibitem{solgas1} V. E. Zakharov, ``Kinetic equation for solitons", Sov. Phys. JETP {\bf 33}, 538 (1971).

\bibitem{solgas2} G. A. El, ``The thermodynamic limit of the Whitham equations", Phys. Lett. A {\bf 311}, 374 (2003).

\bibitem{solgas3} G. A. El and A. M. Kamchatnov, ``Kinetic Equation for a Dense Soliton Gas", Phys. Rev. Lett. {\bf 95}, 204101 (2005).

\bibitem{solgas4} G. A. El, A. M. Kamchatnov, M. V. Pavlov and S. A. Zykov, ``Kinetic equation for a soliton gas and its hydrodynamic reductions", J. Nonlin. Science {\bf 21}, 151 (2011).

\bibitem{dyc} B. Doyon, T. Yoshimura and J.-S. Caux, ``Soliton gases and generalized hydrodynamics", preprint {\tt arXiv:1704.05482} (2017).
 \bibitem{spin1} A. De Luca, M. Collura and J. De Nardis,
	``Non-equilibrium spin transport in the XXZ chain:
persistent currents and emergence of magnetic domains'', Phys. Rev. B (2017), preprint {\tt arXiv:1612.07265}.
\bibitem{spin2} E. Ilievski and J. De Nardis, ``On the microscopic
	origin of ideal conductivity'', preprint {\tt arXiv:1702.02930} (2017).
\bibitem{spin3} V. B. Bulchandani, R. Vasseur, C. Karrasch and
	J. E. Moore, ``Bethe-Boltzmann hydrodynamics and spin transport
	in the XXZ chain'', preprint {\tt arXiv:1702.06146} (2017).
\bibitem{ds} B. Doyon and H. Spohn, ``Dynamics of hard rods with initial
	domain wall state'', J. Stat. Mech. {\bf 2017}, 073210 (2017).
\bibitem{bvkm2} V. B. Bulchandani, R. Vasseur, C. Karrasch and J. E. Moore, ``Solvable hydrodynamics of quantum integrable systems", preprint {\tt arXiv:1704.03466} (2017).
\bibitem{ddky} B. Doyon, J. Dubail, R. M. Konik and T. Yoshimura, ``Large-scale description of interacting one-dimensional Bose gases: generalized hydrodynamics supersedes conventional hydrodynamics", preprint {\tt arXiv:1704.04151} (2017).


\bibitem{Bu17} V. B. Bulchandani, ``On classical integrability of the hydrodynamics of quantum integrable systems", preprint {\tt arXiv:1706.06278} (2017).


\bibitem{prosendiff} M. Ljubotina, M. Znidaric and T. Prosen, ``Spin diffusion from an inhomogeneous quench in an integrable system'', Nat. Comm. {\bf 8}, 16117 (2017).
\bibitem{bonnes} L. Bonnes, F. H. L. Essler and A. M. L\"auchli,
	``Light-cone, dynamics after quantum quenches in spin chains",
	Phys. Rev. Lett. {\bf 113}, 187203 (2014).
\bibitem{nonh} M. E. Tuckerman, Y. Liu, G. Ciccotti and G. J. Martyna,  ``Non-Hamiltonian molecular dynamics: generalizing
Hamiltonian phase space principles to non-Hamiltonian systems'',
	J. Chem. Phys. {\bf 115} 1678 (2001).
\bibitem{dubail} J. Dubail, J.-M. Stephan, J. Viti and P. Calabrese, ``Conformal field theory for inhomogeneous one-dimensional quantum
	systems: the example of non-interacting Fermi gases", SciPost
	Phys. {\bf 2}, 002 (2017).

\end{thebibliography}
\end{document}